\documentclass[preprint2]{aastex}

\shorttitle{Strong Lensing and Dynamic Galactic Mass}
\shortauthors{Guimar\~aes \& Sodr\'e Jr.}

\begin{document}

\title{Bayesian analysis of joint strong gravitational lensing and dynamic galactic mass in SLACS: evidence of line-of-sight contamination}

\author{Antonio C. C. Guimar\~aes and Laerte Sodr\'e Jr.}
\affil{Departamento de Astronomia, Universidade de S\~ao Paulo, \\ Rua do Mat\~ao 1226,CEP 05508-900 S\~ao Paulo - SP, Brazil}
\email{aguimaraes@astro.iag.usp.br}

\begin{abstract}
We readdress the calculation of the mass of early-type galaxies using strong gravitational lensing and stellar dynamics. Our sample comprises 27 galaxies in the Sloan Lens ACS (SLACS) Survey. Comparing the mass estimates from these two independent methods in a Bayesian framework, we find evidence of significant line-of-sight mass contamination. Assuming a power-law mass distribution, the best fit density profile is given by $\rho \propto r^{-1.69\pm0.05}$. We show that neglecting the line-of-sight mass contamination produces an overestimate of the mass attributed to the lens-galaxy by the lensing method, which introduces a bias in favor of a SIS profile when using the joint lensing and dynamic analysis to determine the slope of the density profile. We suggest that the line-of-sight contamination could also be important for other astrophysical and cosmological uses of joint lensing and dynamical measurements.
\end{abstract}


\section{Introduction}

Recently the Sloan Lens ACS (SLACS) Survey \citep{2006ApJ...638..703B} obtained images of dozens galaxy-scale strong gravitational lenses for previously identified early-type galaxies from the Sloan Digital Sky Survey (SDSS). The data of both surveys allow the estimate of the mass of the lens-galaxies by two independent methods.

The mass of a galaxy is one of its most basic properties, nevertheless we are only able to estimate this quantity based on indirect methods. 
There are several ways to estimate the mass of galaxies \citep{1979ARA&A..17..135F,1984ApJ...281L..59T,1996ApJ...473L..17D,2001ApJ...551..643S,2001ApJ...555..572W,2002MNRAS.335..311G,2003ApJ...598..260P,2005ApJ...623L...5F,2006ApJ...648..826L,2006A&A...458..717R}, and different methods are based on different assumptions and measurements, therefore it is of great relevance to test and compare them. Doing so, we can learn not only about the galactic mass, but also about the methods and their assumptions.

One of the main motivations of galactic mass measurement is to know how it is distributed. Some dark matter simulations indicate the emergence of a universal density profile, which describe halos inside a broad mass range of several orders of magnitude, from galactic to cluster scales \citep{1996ApJ...462..563N,1997ApJ...490..493N}. Simulations that include baryons indicate a distinct distribution for the stellar and dark matter components inside a galaxy \citep{2007MNRAS.376...39O,2007arXiv0705.0682B}. In fact, observations confirm a predominance of stellar matter close to the center of galaxies, and a prevalence of dark matter on their outskirts \citep{2005ApJ...623L...5F,2006A&A...458..717R,2006ApJ...648..826L}.

It is of great astrophysical and cosmological interest to know exactly what is the density profile of galaxies, because it can validate or falsify galaxy and structure formation models, and even contain information about the fundamental properties of dark matter \citep{2007astro.ph..1134S}.

In this work we explore the two independent galaxy mass estimates obtained using the strong gravitational lensing effect and stellar dynamics. Both methods are based on very well established theories and measurements, and can be combined to constrain lens models and the astrophysical and cosmological assumptions made. For a theoretical foundation on the subjects one can consult \cite{2006glsw.conf.....M} and \cite{binney87}. We are particularly interested in investigating the role that a possible line-of-sight mass contamination can have on the lensing mass estimate, and therefore on the lens-galaxy density profile inferred from the joint lensing and dynamical analysis.

Work on joint analysis of gravitational lensing and stellar velocity dispersion in early-type lens galaxies started with few single objects \citep{2002ApJ...568L...5K,2004ApJ...611..739T,2005astro.ph..7056H}, but is rapidly accumulating events that now allow statistical treatment.
\cite{treu06} studied the distribution of lens galaxies in the fundamental plane [also treated by \cite{2007astro.ph..1706B}] and infered that within the Einstein radius the isothermal density profile is a good approximation for the SLACS lenses. This last result was also obtained by \cite{2006ApJ...649..599K}, who quantified the variance in the power-law index of the lenses density profile, calculates their dark matter fraction and found no significant evolution of the total inner density slope. \cite{2007astro.ph..1589G} used weak gravitational lensing measurements around lens galaxies that are strong lenses to investigate the density profile at larger radii, finding again an isothermal mass distribution for SLACS lenses.

Data from SLACS is especially suitable for joint strong lensing and dynamic analysis because it allows precise determination of the Einstein radius for each lens-galaxy of a relatively homogeneous sample of early-type galaxies. And, at the same time, SDSS has precise stellar velocity dispersion measurements for all lenses, and redshifts of all objects. Previously, the main database for strong gravitational lenses containing the same kind of data was CASTLES, the CfA-Arizona Space Telescope Lens Survey \citep{1999astro.ph.10025F}, however in lower number and homogeneity.

The outline of the paper is the following. In Section \ref{data} we present and explain the data used. In Section \ref{MassEstimate} we introduce the theoretical framework of mass estimate by strong gravitational lensing and by stellar dynamics, and present some models for the line-of-sight mass contamination and density profile. In Section \ref{bayes} we lay down the likelihood and Bayes analysis formalism used to quantify the goodness of fit and compare the various models considered. In Section \ref{results} we present our results, which are discussed in Section \ref{conclusion} together with our conclusions.

\section{The Data}
\label{data}

We use data compiled from \cite{2006ApJ...649..599K} and \cite{2007astro.ph..1589G}, constructing a sample of 27 strong gravitational lensing events where the lenses are isolated early-type galaxies (E+S0). 

The selected set of galaxies is part of the SLACS Survey \citep{2006ApJ...638..703B,treu06}, which is a Hubble Space Telescope (HST) Snapshot imaging survey for strong gravitational galactic lenses. The candidates for the HST imaging were selected spectroscopically from the SDSS database and are a sub-sample of the SDSS Luminous Red Galaxy (LRG) sample.

For each lens system we are interested in the redshift of the background lensed source $z_s$, the redshift of the lens $z_l$, the average stellar velocity dispersion inside an aperture $\sigma_{ap}$ and the Einstein radius $\theta_E$.
See Table \ref{datatable} for the compiled data set.

The source and lens redshifts were determined from the SDSS spectra, and the stellar velocity dispersion corresponds to the light-weighted average inside the 3\arcsec  diameter SDSS fiber. 

The Einstein radii were determined from HST images using strong lensing modeling of the lenses and reconstruction of the unlensed sources \citep{2006ApJ...649..599K,2007astro.ph..1589G}. The uncertainties on $\theta_E$ were reported to be around 5\%, so we use this value for all Einstein radii.
We note that the Einstein radius determined from this procedure is a robust attribute of the lens, being little sensitive to the lens model used [see Kochanek contribution in \cite{2006glsw.conf.....M}].

\section{Galaxy Mass Estimate} 
\label{MassEstimate}

We briefly review the independent mass estimate methods based on strong gravitational lensing and stellar dynamics, assuming spherical symmetry.

\subsection{Strong Gravitational Lensing}

When a background source is aligned  with an intervening mass concentration (lens) the source light can be strongly deflected and form a luminous ring around the lens (Einstein ring). Less well aligned sources can generate arcs and multiple images. These tracers of the tangential critical curve can be used to estimate the mass enclosed within the Einstein radius $R_E=\theta_E D_L$,
\begin{equation}
M_L = \frac{c^2}{4G}\frac{D_L D_S}{D_{LS}} \theta_E^2 \;, \label{M_L}
\end{equation}
where $D_{[L,S,LS]}$ is the angular-diameter distance of the lens, source, and between lens and source, respectively.

Note that the source light ray is deflected not only by the mass associated to the lens galaxy, but by all mass along the line-of-sight, generating a contamination that, if ignored, can imply in an overestimate of the lens mass.

We suggest some simple models to subtract the light-of-sight mass contamination from the lens galaxy mass estimate.
We adopt a notation in which $M_L$ is the mass estimate from Eq.(\ref{M_L}) and 
$m_L$ is the lens mass estimate taking into account the line-of-sight mass contamination.
In the first model we subtract a constant contamination from the mass estimate of each lens and  
$m_L=M_L-p$, where $p$ is a free parameter. In the second model we assume a constant surface density contamination, so, for each lens, the mass overestimate is proportional to the lens area,  
$m_L=M_L-pA$.
In our third model we propose that the line-of-sight mass contamination is not only proportional to the lens area, but also to the lens mass, $m_L=M_L-pM_LA$. This is an attempt to see if more massive lenses have a higher line-of-sight mass contamination (this could be the case if they are tracing higher density regions).

The possible dependence of the line-of-sight mass contamination with the redshift of the source is contemplated by the models $m_L=M_L-pz_L$ and $m_L=M_L-pz_LA$, where the latter roughly mimics the volume covered by the lensed light rays from the galaxy lens.

\subsection{Stellar Dynamics}

From Jeans equations \citep{binney87} for a spheric halo we have that for any isotropic halo profile, the radial stellar velocity dispersion is given by
\begin{equation}
\sigma^2(R,z)=\frac 1{\rho} \int_z^\infty \rho \frac{\partial \Phi}{\partial z} dz \;,\label{vel-disp}
\end{equation}
where we assumed a constant mass to light ratio, and $\Phi$ is the gravitational potential. 

We define the average line-of-sight velocity dispersion within an aperture of radius $R$ as 
\begin{equation}
\langle \sigma^2_\shortparallel \rangle(<R) \equiv \frac {\int_C \rho \sigma^2 dV}
{\int_C \rho dV} \; , \label{sigma2}
\end{equation}
where the volumetric mean of the velocity dispersion is calculated inside an infinite cylinder $C$ with axis along the line-of-sight, weighted by the density profile.

If we now assume a halo profile of the form $\rho \propto r^\gamma$ and call $\sigma^2_{ap}=\langle\sigma^2_\shortparallel\rangle(<R_{ap})$ the velocity dispersion measured in an aperture of radius $R_{ap}$, then the mass inside the cylinder $C$ [denominator in the right side of Eq. (\ref{sigma2})] is
\begin{equation}
M_D(<R) = \frac \pi{G} \sigma^2_{ap} R \left(\frac R{R_{ap}}\right)^{2+\gamma}f(\gamma) \; , \label{dynamic_mass}
\end{equation}
where 
\begin{equation}
f(\gamma) = - \frac 1{\sqrt{\pi}} \frac{(5+2\gamma)(1+\gamma)}{(3+\gamma)}
\frac{\Gamma(-\gamma-1)}{\Gamma(-\gamma-\frac 3{2})} 
\left[\frac{\Gamma(-\frac \gamma{2}-\frac 1{2})}{\Gamma(-\frac \gamma{2})} \right]^2 .
\end{equation}
\cite{2006EAS....20..161K} found similar expressions.

For $\gamma=-2$ (singular isothermal sphere, or SIS), Eq.(\ref{dynamic_mass}) reduces to the well known expression
\begin{equation}
M_D^{SIS}(<R) = \frac \pi{G} \sigma^2_{ap} R \;.
\end{equation}
Fig. \ref{fac_p_law} shows the behavior of the factors present in Eq. (\ref{dynamic_mass}) that depend on $\gamma$. For $R=R_{ap}$ the SIS profile maximizes the mass contained in a cylinder for a fixed average line-of-sight velocity dispersion.

SLACS data have a velocity dispersion measured inside a $3\arcsec$ diameter aperture ($\theta_{ap}=1.5\arcsec$) and its average Einstein radius is $\langle \theta_E \rangle=1.2\arcsec$, which imply an average $R_E/R_{ap}=0.8$. 

Therefore we have two models for the mass enclosed inside the Einstein radius estimated from the dynamical method, $m_D$. 
The first model considers only a SIS density profile and there is no free parameter, $m_D=M_D^{SIS}(<R_E)$.  The second model allows a free index for the power-law profile, $m_D=M_D(<R_E,\gamma)$.

\section{Bayesian Analysis}
\label{bayes}

Since we raise the possibility of various models for the dynamic mass estimate and line-of-sight contamination, we need a formalism to compare these models. 
So, in addition to a likelihood analysis, we also calculate the Bayesian Information Criteria ($BIC$) and the Bayes evidence \citep{2007MNRAS.tmpL..34L}.

The likelihood of both estimate methods to give the same galaxy mass is
\begin{equation}
{\cal L}_i = \frac 1 {\sqrt{2\pi (\sigma^2_{L,i}+\sigma^2_{D,i})}} \exp \left[-\frac{(m_{L,i}-m_{D,i})^2}{2(\sigma^2_{L,i}+\sigma^2_{D,i}) } \right] ,
\end{equation}
and the likelihood for a whole set of $N$ galaxies is given by
\begin{equation}
{\cal L} = \prod _{i=1}^N  {\cal L}_i .
\end{equation}

The Bayesian Information Criteria ($BIC$) is defined as
\begin{equation}
BIC \equiv -2\ln{\cal L}_{max} + k \ln N\; ,
\end{equation}
where ${\cal L}_{max}$ is the maximum likelihood, $k$ is the number of parameters in the model, and $N$ is number of data points.

The Bayesian evidence is defined by
\begin{equation}
E \equiv \int {\cal L}(p) P(p) dp \; ,
\end{equation}
where $P(p)$ is the prior probability distribution for parameter $p$. We use a flat prior probability distribution for all the parameters tested.

To compare two models we calculate the difference between their Bayesian Information $\Delta BIC$ and $\Delta \ln E$. 
A $\Delta BIC>2$ indicates positive evidence in favor of the model with lower $BIC$, and  $\Delta BIC>6$ indicates strong evidence.
Under examination of the Bayesian evidences, $\Delta \ln E<1$ does not constitute evidence, $1<\Delta \ln E<2.5$ means a significant evidence, $2.5<\Delta \ln E<5$ means strong evidence, and $5<\Delta \ln E$ means decisive evidence in favor of the model with higher $E$ \citep{jeffreys1961}.

The Bayesian evidence is a more rigorous and general criteria than the $BIC$ \citep{2004MNRAS.351L..49L}, but we include the latter because it is much simpler to calculate, so it is interesting to see if both criteria agree.

We note that \cite{2007astro.ph..1372B} developed a framework for joint gravitational lensing and stellar dynamics analysis using Bayesian statistics, but which is very distinct from our approach.

\section{Results}
\label{results}

We use the strong gravitational lensing and stellar dynamics theories to compute independent mass estimates inside the Einstein radius for the 27 selected galaxies from SLACS.
We adopt a concordance $\Lambda$CDM model ($\Omega_m=0.3$ and $\Omega_\Lambda=0.7)$) to provide the redshift-distance relation needed for the calculations.

\subsection{SIS Density Profile}

\begin{figure*}
\epsscale{2.2}
\plottwo{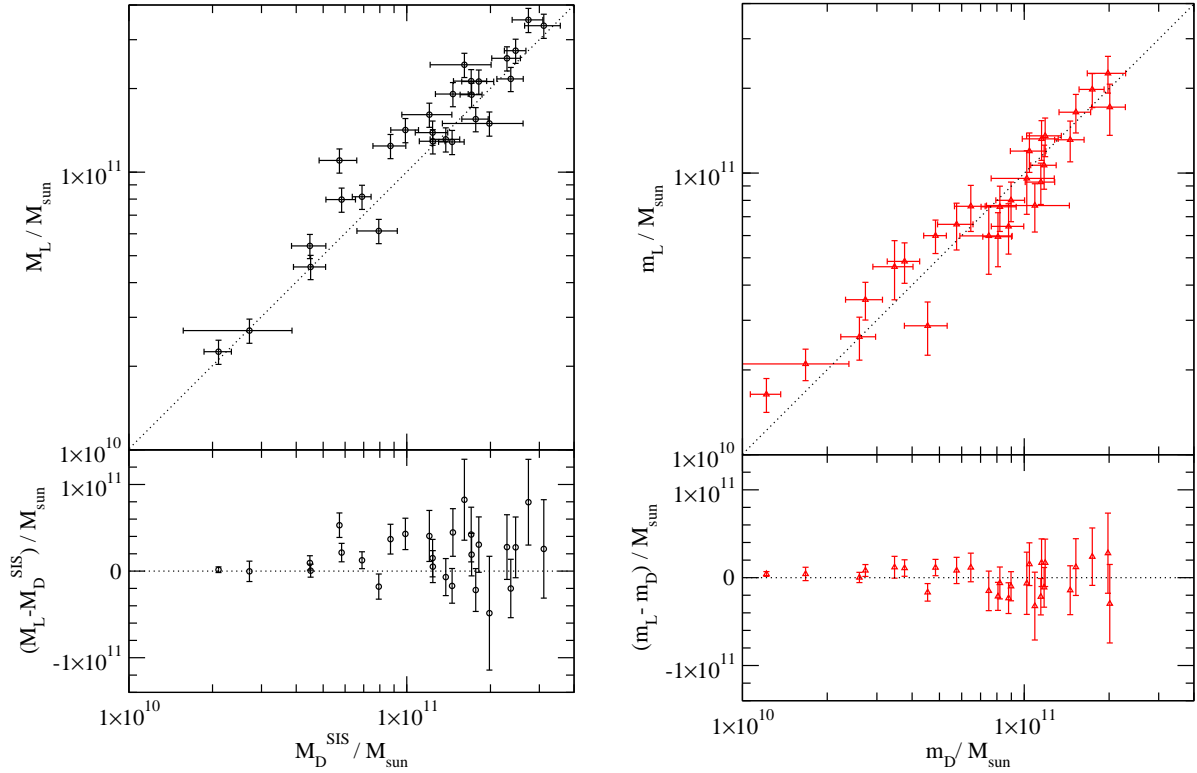}{f2b.eps}
\caption{Dynamic and lensing estimates of the mass enclosed by the Einstein radius. The left panel shows the results for our simplest model, a SIS with no line-of-sight mass contamination. The right panel is for the best power-law density profile model with line-of-sight mass contamination subtraction.
\label{SDSSlenses_masses}}
\end{figure*}

For our simplest model, a SIS density profile with no line-of-sight contamination, we find that the lensing method overestimates the galaxy mass compared to the dynamic method. 
This is noticeable in the left plot at Fig.\ref{SDSSlenses_masses}. A linear fitting between the two mass estimates gives that $M_L\approx (1.06\pm0.08)M_D^{SIS}$.

One possible explanation for the observed overestimate of the lensing method in relation to the dynamic one is that the former also captures mass contaminants in the line-of-sight, while the dynamic method is only sensitive to the galaxy internal gravitational potential.
We test five models for the line-of-sight contamination (models I1 to I5) that have one free parameter and calculate their likelihood, $BIC$ and Bayes evidence (see Table \ref{models_results}). All models with contamination perform better than the one with no contamination (model I0). They have higher likelihood, lower $BIC$ and higher Bayesian evidence.

The line-of-sight mass contamination is significant, 4\% in model I1, 14\% in model I2, 11\% in model I3, 9\% in model I4 and 12\% in model I5. This is computed having as base an average lensing mass  
$\langle M_L\rangle=1.57\times10^{11}M_\sun$, an average lens area 
$\langle A\rangle=28.1 kpc^2 h^{-2}$ and an average lens redshift
$\langle z_L\rangle=0.222$ for our set of 27 galaxies.

\subsection{Power-Law Density Profile}

If we relax the SIS density profile assumption and use instead a power-law density profile for the galaxies we obtain models with one more free parameter, the power-law index. 

\begin{figure*}
\epsscale{1.0}
\plotone{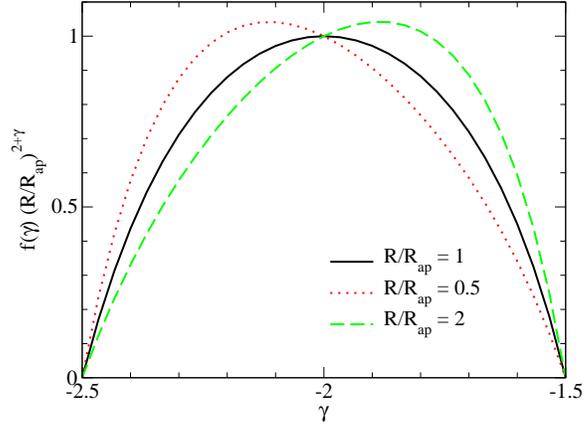}
\caption{The factor present in Eq.(\ref{dynamic_mass}), which generalizes the mass enclosed inside a radius $R$ for a SIS ($\gamma=-2$) to a power-law density profile. \label{fac_p_law}}
\end{figure*}

The model with a power-law profile and no line-of-sight contamination (model P0) has maximum likelihood when its power-law index is $-2.02\pm0.04$, what means that the SIS profile is in fact preferred by the system. 
This occurs because the lensing mass is larger than the dynamic mass and it is an index close to -2, but slightly lower than that, which maximizes the dynamic mass inside the Einstein radius, as is show in Fig.\ref{fac_p_law}. For our galaxy sample $\langle R_E/R_{ap}\rangle=0.8$, so the effective curve for it is something in between the solid curve for $R/R_{ap}=1$ and the dotted curve for $R/R_{ap}=0.5$.

All models with power-law profile (models P0 to P5) have higher likelihoods than their SIS counterparts (models I0 to I5, respectively). That is expected since the added parameter creates space for better fittings, but the extra model complexity is penalized by the $BIC$ and Bayes evidence measures. In fact only for model P2 the gain in the likelihood is enough to give a lower $BIC$ and higher Bayesian evidence than for its SIC counterpart, model I2. This model gives the highest likelihood, lowest $BIC$ and highest Bayes evidence among all models investigated. Therefore it is adopted as referential for comparisons with the other models.
The right plot at Fig.\ref{SDSSlenses_masses} illustrates how much better is the agreement between the two mass estimate methods in model P2. The mass estimate points are closer to the identity line, and there is a lower dispersion. A linear fitting to this data gives that $m_L\approx (0.99\pm0.06)m_D$.

\begin{figure*}
\epsscale{1.0}
\plotone{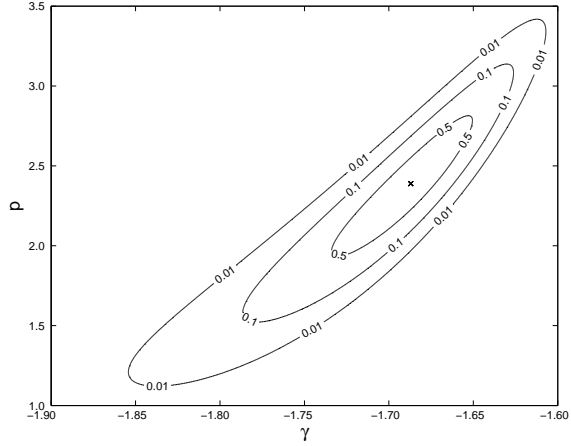}
\caption{Likelihood contours in the parameter space of model P2. Likelihood is in units of ${\cal L}_{max}$ for this model, and $p$ is in units of $10^9M_\sun kpc^{-2}h^2$.} \label{model12_likelihood}
\end{figure*}

At Fig.\ref{model12_likelihood} we show the likelihood of model P2 in its parameter space, the parameter that quantifies the line-of-sight mass contamination and the power-law index of the density profile. There is a degeneracy between the two parameters; if we take a high power-law index (flatter profile) the system demands a higher line-of-sight mass contamination.
In contrast with all other models probed, which prefer a SIS profile, model P2 has maximum likelihood at a higher power-law index $\gamma=-1.69\pm0.05$. As a consequence of this flatter profile, farther from the SIS maximization point discussed before in relation to Fig.\ref{fac_p_law}, a higher line-of-sight mass contamination is necessary to fill the gap between the dynamic mass estimate and the lensing estimate. In fact, model P2 has a considerably higher contaminant surface density than model I2 and its line-of-sight contamination represents 43\% of the average lensing mass.
This is a very high contamination when compared with the ones obtained in the other models, 4\% in model P1, 14\% in models P3 and P4, and 17\% in model P5.

\section{Discussion and Conclusions}
\label{conclusion}

We used two independent methods, one based on strong gravitational lensing and another on stellar dynamic, to estimate the mass of 27 early-type galaxies from SLACS. 
Because each method has different assumptions and dependencies, the comparison of the mass estimate obtained from each method enables us to investigate some underlying aspects of the galaxies and of the lensing system, namely the average galaxy profile and the line-of-sight mass contamination.

Likelihood and Bayesian analysis were employed to compare models and quantify the evidence of our results. 
We find that there is decisive evidence of line-of-sight mass contamination for the early-type galaxies selected. This is significant, considering that the lens sample was constituded by isolated galaxies. Probably because of this, we do not find evidence of a higher line-of-sight mass contamination for more massive lenses, as would be expected if the galaxies belonged to groups or clusters. 
Previously, \cite{2006ApJ...641..169M} and \cite{2006ApJ...646...85W} also found a significant line-of-sight effect on strong lens galaxies.
\cite{2007ApJ...660L..31M} found that even in underdense local enviroments, the line-of-sight contamination may give a considerable contribution to galaxy-scale strong lenses, in agreement with our results.
However, the line-of-sight mass contamination is more strongly associated to the area of the lens than with its redshift (we also tested the dependence with the source redshift and lens-source distance and found none). This is an indicative that the contamination comes from material in the vicinity of the lens and not from along the whole line-of-sight, what is in accordance with the cosmological average of zero convergence in a random line-of-sight due to the canceling of over- and under-dense regions \citep{2002MNRAS.337..631G}.

Our most successful model, which assumes a power-law spherical profile and uniform surface density line-of-sight contamination, gives evidence for a flatter density profile than the one from a singular isothermal sphere. In fact, the power-law index found is $-1.69\pm0.05$, but it implies for this model a very high line-of-sight mass contamination (43\%). Such unexpected high contamination rises suspition on the limitations of the assumptions and models adopted.

The main assumptions that we made concerning the lens galaxies were the sphericity and smoothness of the density profiles, no rotational support, power-law profile and constant mass-to-light ratio along the galaxy radius. These suppositions must be taken as approximations, since there is evidence against each one of them, as we discuss in the sequence.

An ellipsoidal symmetry is more general and reliable than a spherical one for the lens galaxies \citep{2007arXiv0705.0682B,2007MNRAS.377...50H}, but it is also an approximation and incurs in one extra parameter. The existence of substructure on galactic scales may also affect the mass estimates \citep{2006ApJ...643..154Y}, but its inclusion involves a heavy complexity (more parameters). \cite{2002ApJ...574L.129S}, assuming spherical symmetry and using lensing and dynamic analysis, found a flat inner slope for the dark matter distribution in a particular galaxy cluster. This result was questioned by \cite{2005astro.ph..9323M}, who showed the importance of ellipticity and substructures using numerical simulations, 

Contrary to old beliefs, \cite{2007astro.ph..3531E} showed that some early-type galaxies can have a significant rotation component, impling some rotational support that would require a correction of the dynamic mass estimate. The power-law approximation for the overall galactic density profile has been favored by several works, apparently as a result of complementary baryonic and dark matter profiles, also suggesting a shortcoming of a constant mass-to-light ratio approximation \citep{2005astro.ph..7056H,2005ApJ...623L...5F,2006ApJ...648..826L,2007arXiv0705.0682B}.

Concerning the cosmological framework, the model assumed (concordance $\Lambda$CDM) served the unique purpose of providing a redshift-distance relation. If this relation could be obtained independently of cosmic models, then they could be factored out of the analysis. Nevertheless, it is the very redshift-distance relation that serves as basis to existing suggestions of using joint lensing and dynamic analysis to probe cosmological models and their parameters. 
In a era of precision cosmology, the proper accounting of line-of-sight effects would be fundamental in these studies.

It would be interesting to relax some of our assumption and investigate the role of the extra parameters. However, models with a large number of parameters may provide higher likelihood, but lower Bayesian evidence, since the added complexity must pay its price in a Bayesian sense. A larger galaxy sample would probably be necessary to reduce the increased degeneracies among the free parameters.

The perspective of more data of the kind necessary for joint lensing and dynamic studies is encouraging, as well as the potentiality of using them to investigate galaxy properties and cosmological parameters. On the galactic scale, the density profile, shape and substructure are the most interesting targets, and on the cosmological scale the redshift-distance relation. Nevertheless, our work shows that it is essential for these studies to consider the line-of-sight mass contamination, since neglecting it contribute to an overestimate of the galactic mass by the strong lensing method, what artificially forces the system to prefer a SIS profile because that maximizes the dynamic mass estimate.

\acknowledgments
The authors thank CNPq and FAPESP for financial support and the SLACS and SDSS teams for the databases used in this work. Funding for the SDSS has been provided by the Alfred P. Sloan Foundation, the Participating Institutions, the National Aeronautics and Space Administration, the National Science Foundation, the US Department of Energy, the Japanese Monbukagakusho, and the Max Planck Society. The SDSS is managed by the Astrophysical Research Consortium (ARC) for the Participating Institutions. The Participating Institutions are The University of Chicago, Fermilab, the Institute for Advanced Study, the Japan Participation Group, The Johns Hopkins University, the Korean Scientist Group, Los Alamos National Laboratory, the Max-Planck-Institute for Astronomy (MPIA), the Max-Planck-Institute for Astrophysics (MPA), New Mexico State University, University of Pittsburgh, University of Portsmouth, Princeton University, the United States Naval Observatory, and the University of Washington.


\begin{table*}
\begin{center}
\caption{Compiled data from \cite{2006ApJ...649..599K} (indicated with a * after the name) and \cite{2007astro.ph..1589G} (some of which were already present in the former paper).  \label{datatable}}
\begin{tabular}{lllll}
\tableline\tableline
Name   & $z_{l}$ & $z_s$ & $\theta_E(\arcsec)$ & $\sigma_{ap}(km\,s^{-1})$ \\
\tableline
SDSS J002907.8-005550 & 0.227  & 0.931  & 0.82  & $228\pm18$    \\
SDSS J0037-0942 *     &	0.195  & 0.632  & 1.47	& $265\pm10$	\\	
SDSS J015758.9-005626 & 0.513  & 0.924	& 0.72	& $295\pm 7$	\\
SDSS J021652.5-081345 & 0.332  & 0.523	& 1.15	& $333\pm23$	\\
SDSS J025245.2+003958 & 0.280  & 0.982	& 0.98	& $164\pm12$	\\	
SDSS J033012.1-002052 & 0.351  & 1.107	& 1.06	& $212\pm21$	\\	
SDSS J072805.0+383526 & 0.206  & 0.688	& 1.25	& $214\pm11$	\\	
SDSS J0737+3216 *     &	0.322  & 0.581  & 1.03	& $310\pm15$	\\	
SDSS J080858.8+470639 & 0.219  & 1.025	& 1.23	& $236\pm11$	\\	
SDSS J090315.2+411609 & 0.430  & 1.065	& 1.13	& $223\pm27$	\\	
SDSS J091205.3+002901 & 0.164  & 0.324	& 1.61	& $326\pm12$	\\	
SDSS J0956+5100 *     &	0.240  & 0.470  & 1.32	& $299\pm16$	\\	
SDSS J095944.1+041017 &	0.126  & 0.535 	& 1.00	& $197\pm13$	\\	
SDSS J102332.3+423002 & 0.191  & 0.696	& 1.30 	& $242\pm15$	\\	
SDSS J110308.2+532228 & 0.158  & 0.735	& 0.84	& $196\pm12$	\\	
SDSS J120540.4+491029 & 0.215  & 0.481	& 1.04	& $280\pm13$	\\	
SDSS J125028.3+052349 & 0.232  & 0.795	& 1.15	& $252\pm14$	\\	
SDSS J1330-0148 *     &	0.081  & 0.711	& 0.85	& $178\pm 9$	\\	
SDSS J140228.1+632133 & 0.205  & 0.481	& 1.39	& $267\pm17$	\\	
SDSS J142015.9+601915 & 0.063  & 0.535	& 1.04	& $205\pm43$	\\	
SDSS J162746.5-005358 & 0.208  & 0.524 	& 1.21	& $290\pm14$	\\	
SDSS J163028.2+452036 & 0.248  & 0.793 	& 1.81	& $276\pm16$	\\	
SDSS J223840.2-075456 & 0.137  & 0.713	& 1.20	& $198\pm11$	\\	
SDSS J230053.2+002238 & 0.228  & 0.463	& 1.25	& $279\pm17$	\\	
SDSS J230321.7+142218 & 0.155  & 0.517	& 1.64	& $255\pm16$	\\	
SDSS J2321-0939 *     &	0.082  & 0.532  & 1.57	& $236\pm 7$	\\	
SDSS J234111.6+000019 & 0.186  & 0.807	& 1.28	& $207\pm13$	\\	
\tableline
\end{tabular}
\end{center}
\end{table*}

\begin{table*}
\begin{center}
\caption{Comparison among models and best parameters. 
\label{models_results}}
\small
\begin{tabular}{clccccccc}
\tableline\tableline
Model & $m_L$ & $p_{\sup}$ & $p_{best}$ & $\gamma_{best}$ & ${\cal L}_{max}$ & $\Delta BIC$ & $\Delta \ln E$ \\
\tableline
I0 & $M_L$  & -  & - & - & 2.89$\cdot 10^{-301}$ & 33.7 & 14.4 \\
I1 & $M_L-p$  & $10^{11}$ & $(7\pm2)\cdot10^9M_\sun$ & - & 2.46$\cdot10^{-299}$ & 28.1  & 12.8 \\
I2 & $M_L-pA$  & 2$\cdot10^9$ & $(8\pm2)\cdot10^8M_\sun kpc^{-2}h^2$ & - & 4.07$\cdot10^{-296}$ & 13.3 & 4.1 \\
I3 & $M_L-pM_LA$  & 3$\cdot10^{-2}$ & $(4.1\pm0.9)\cdot10^{-3}kpc^{-2}h^2$ & - & 6.95$\cdot10^{-297}$ & 16.9 & 6.9 \\
I4 & $M_L-pz_L$  & 5$\cdot10^{11}$ & $(7\pm2)\cdot10^9M_\sun$ & - & 5.82$\cdot10^{-298}$ & 21.8 & 9.2 \\
I5 & $M_L-pz_LA$  & 1$\cdot10^{10}$ & $(3.1\pm0.7)\cdot10^9M_\sun kpc^{-2}h^2$ & - & 1.04$\cdot10^{-296}$ & 16.0 & 5.6 \\
P0 & $M_L$   & - & - & $-2.02\pm0.04$ & 5.15$\cdot10^{-301}$ & 35.9 & 16.2 \\
P1 & $M_L-p$   & $10^{11}$ & $(7\pm3)\cdot10^9M_\sun$ & $-1.98\pm0.05$ & 3.51$\cdot10^{-299}$ & 30.7 & 14.4 \\
P2 & $M_L-pA$ & $10^{10}$ &  $(2.4\pm0.4)\cdot10^9M_\sun kpc^{-2}h^2$ & $-1.69\pm0.05$ & 1.61$\cdot10^{-292}$ & 0 & 0 \\
P3 & $M_L-pM_LA$  & 3$\cdot10^{-2}$ & $(5\pm1)\cdot10^{-3}kpc^{-2}h^2$ & $-1.9\pm0.1$ & 8.97$\cdot10^{-297}$ & 19.6 & 8.0 \\
P4 & $M_L-pz_L$   & $5\cdot10^{11}$ & $(1.0\pm0.3)\cdot10^{11}M_\sun$ & $-1.87\pm0.07$ & 2.40$\cdot10^{-297}$ & 22.3 & 5.1 \\
P5 & $M_L-pz_LA$ & $10^{10}$ &  $(4.2\pm1.0)\cdot10^9M_\sun kpc^{-2}h^2$ & $-1.88\pm0.08$ & 3.46$\cdot10^{-296}$ & 16.9 & 6.3 \\
\tableline
\end{tabular}
\normalsize
\tablecomments{For models I0-5, $m_D=M_D^{SIS}$, and for models P0-5, $m_D=M_D(\gamma)$. The lower limit of the flat prior for the line-of-sight contamination is 0 and the superior limit is $p_{\sup}$, $p_{best}$ is the best fit value (maximum likelihood). The power-law index for the density profile has a flat prior $-2.5<\gamma<-1.5$ and best fit value  $\gamma_{best}$. ${\cal L}_{max}$ is the maximum likelihood, $\Delta BIC$ the difference in the Bayesian Information Criteria, and $\Delta \ln E$ the Bayesian Evidence ratio, both computed in relation to model P2, which has $BIC=1350.3$ and $E=5.19\cdot10^{-295}$.}
\end{center}
\end{table*}

\end{document}